\documentclass[useAMS,usenatbib]{mn2e}
\usepackage{times}
\usepackage{graphicx}
\title[ALMA imaging of SDP.81 -- I.]{ALMA imaging of SDP.81 -- I. A pixelated reconstruction of the far-infrared continuum emission}
\author[M. Rybak et al.]{M. Rybak,$^{1}$\thanks{E-mail: rybak@mpa-garching.mpg.de} J. P. McKean,$^{2,3}$ S. Vegetti,$^{1}$ P. Andreani$^{4}$ and S. D. M. White$^{1}$\\
$^{1}$Max Planck Institute for Astrophysics, Karl-Schwarzschild-Strasse 1, D-85740 Garching\\
$^{2}$Netherlands Institute for Radio Astronomy (ASTRON), P.O. Box 2, 7990 AA Dwingeloo, The Netherlands\\
$^{3}$Kapteyn Astronomical Institute, University of Groningen, P.O. Box 800, 9700 AV Groningen, The Netherlands\\
$^{4}$European Southern Observatory, Karl-Schwarzschild-Strasse 2, 85748, Garching, Germany}

\begin{document}

\date{Accepted 2015 April 15.  Received 2015 April 8; in original form 2015 March 6}

\pagerange{\pageref{firstpage}--\pageref{lastpage}} \pubyear{2015}

\maketitle

\label{firstpage}

\begin{abstract}
We present a sub-50 pc-scale analysis of the gravitational lens system SDP.81 at redshift 3.042 using Atacama Large Millimetre/submillimetre Array (ALMA) science verification data. We model both the mass distribution of the gravitational lensing galaxy and the pixelated surface brightness distribution of the background source using a novel Bayesian technique that fits the data directly in visibility space. We find the 1 and 1.3~mm dust emission to be magnified by a factor of $\mu_{\rm tot} = $~17.6\,$\pm$\,0.4, giving an intrinsic total star-formation rate of 315\,$\pm$\,60~M$_\odot$~yr$^{-1}$ and a dust mass of 6.4\,$\pm$\,1.5\,$\times$\,10$^{8}$~M$_{\odot}$. The reconstructed dust emission is found to be non-uniform, but composed of multiple regions that are heated by both diffuse and strongly clumped star-formation. The highest surface brightness region is a $\sim$1.9$\,\times$\,0.7~kpc disk-like structure, whose small extent is consistent with a potential size-bias in gravitationally lensed starbursts. Although surrounded by extended star formation, with a density of 20--30\,$\pm$\,10~M$_\odot$~yr$^{-1}$~kpc$^{-2}$, the disk contains three compact regions with densities that peak between 120--190\,$\pm$\,20~M$_\odot$~yr$^{-1}$~kpc$^{-2}$. Such star-formation rate densities are below what is expected for Eddington-limited star-formation by a radiation pressure supported starburst. There is also a tentative variation in the spectral slope of the different star-forming regions, which is likely due to a change in the dust temperature and/or opacity across the source. 
\end{abstract}


\section{Introduction}

\begin{figure}
\begin{centering}
\begin{tabular}{ll}
\hspace{-0.5cm}
\vspace{-0.3cm}
\includegraphics[scale=0.18]{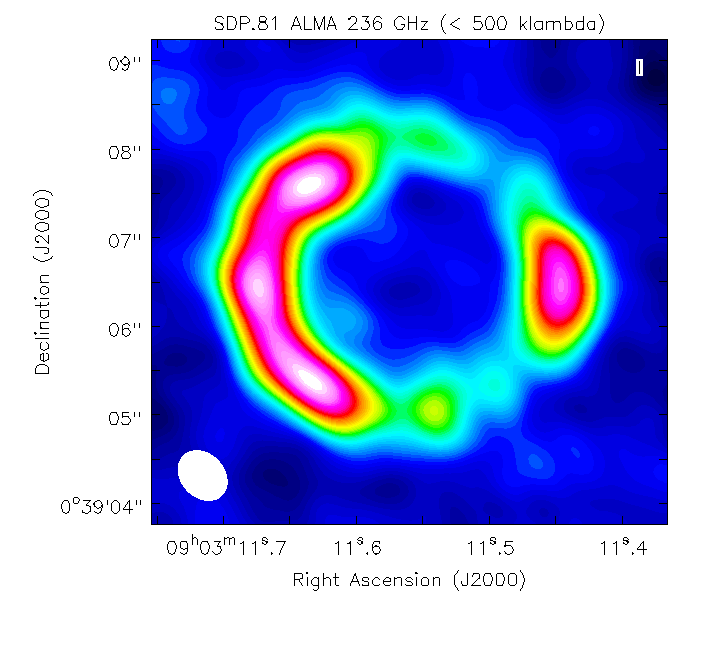}&\hspace{-0.7cm}
\includegraphics[scale=0.18]{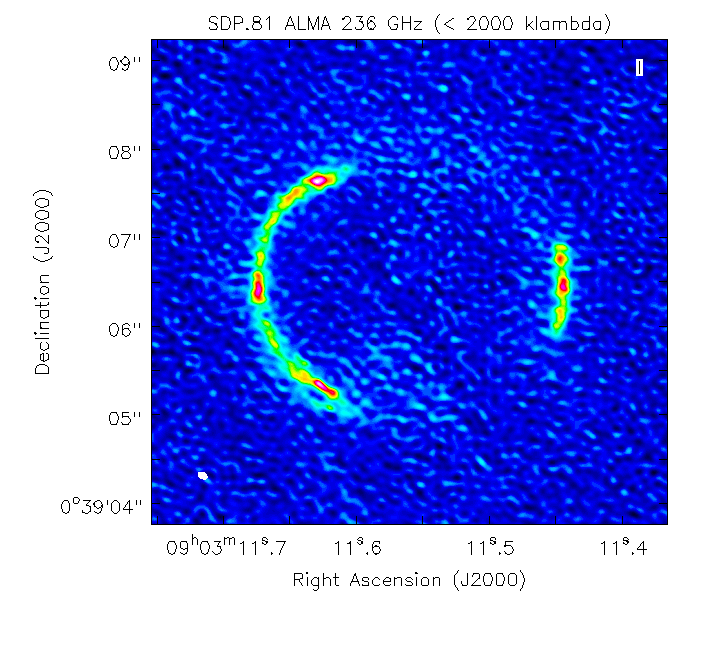}\\
\hspace{-0.5cm}
\vspace{-0.5cm}
\includegraphics[scale=0.18]{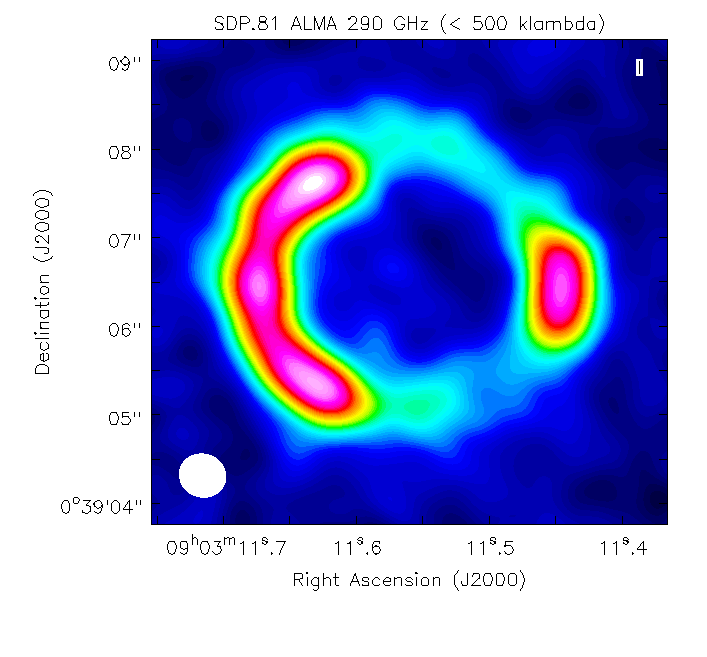}&\hspace{-0.7cm}
\includegraphics[scale=0.18]{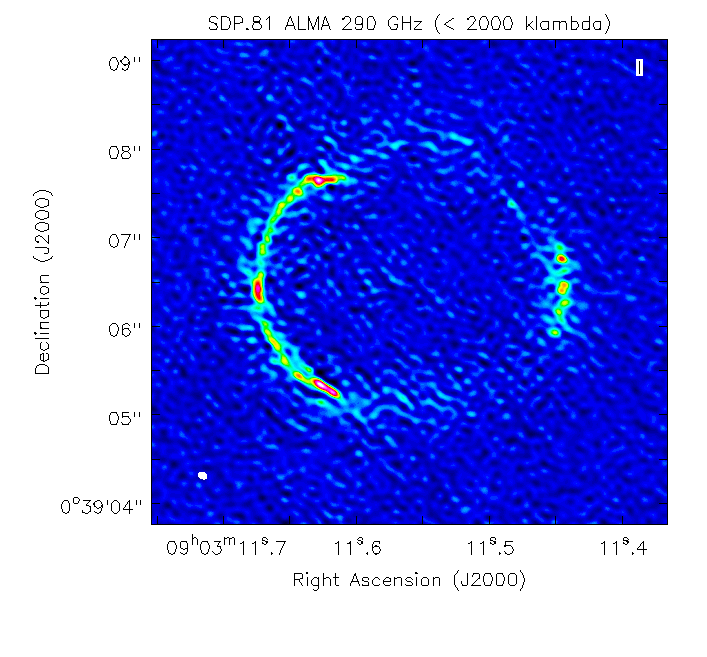}
\end{tabular}
\end{centering}
\caption{The ALMA continuum imaging of SDP.81 at 236 GHz (Upper) and 290 GHz (Lower). The images have been produced at two angular resolutions using a cut in the {\it uv}-data of 0.5 (left) and 2~M$\lambda$ (right), and weightings (natural and Briggs robust = 0) to emphasise the structure seen on different scales. The synthesised restoring beam is shown as the white ellipse in the bottom left corner of each map. These maps are only shown to illustrate the surface brightness distribution of the source. It is the visibility data that are used for the gravitational lens modelling and source reconstruction.}
\label{fig:images}
\end{figure}

Submillimetre (sub-mm) galaxies are understood to be the main drivers of star-formation activity at high redshift and are believed to be progenitors of present-day early-type galaxies (e.g. \citealt{blain02,Swinbank06}). Their large far-infrared luminosities ($\geq$\,10$^{12}$~L$_{\odot}$) are due to stellar ultraviolet (UV) radiation being absorbed and re-emitted at far-infrared wavelengths within dusty star-forming regions. Thus, studies of the dust emission at (rest-frame) far-infrared wavelengths can be used to trace star-formation activity directly, even if the stars themselves suffer from heavy extinction due to the dust \citep{Kennicutt1998}.

While detailed studies of the star-formation processes within these objects have not been possible due to the arcsecond-scale angular resolution of the early instruments (e.g. SCUBA; MAMBO), the advent of interferometric arrays, such as the Atacama Large Millimetre/submillimetre Array (ALMA) and the Submillimetre Array (SMA), has led to a significant improvement in our knowledge about the structure of starbursting galaxies at these wavelengths \citep{Karim2013}. Furthermore, the study of gravitationally lensed sub-mm galaxies has allowed an investigation of the formation of high redshift galaxies at a resolution comparable to what has already been achieved at lower redshift \citep{swinbank10}. Due to the combination of the magnification provided by the foreground gravitational lenses, and the fact that the observed flux density of sub-mm sources is roughly constant across a wide range of redshifts (the so-called {\it negative K-correction}), distant sources up to a redshift of $z$\,$\sim$\,5.6 have been detected in recent sub-mm surveys \citep{Negrello2010,Bussmann2012,Vieira13}.

It is now possible to study the detailed star-formation processes within these galaxies at $<$100~mas angular resolution with the long baselines provided by ALMA. Such studies will allow the size and structure of the heated dust emission to be directly observed for the first time, which is important for determining the actual magnification of the emission regions without the effects of differential magnification. This can establish whether there is a size-bias in gravitationally lensed starbursting galaxies and place robust limits on the star-formation rate density within these galaxies.

In this letter, we present the analysis of high angular resolution imaging of the heated dust continuum emission from the gravitationally lensed starburst galaxy SDP.81 (H-ATLAS J090311.6+003906; $z =$ 3.042\,$\pm$\,0.001; \citealt{Negrello2010,Frayer2011,Bussmann2013,Dye14,Negrello14}) that were taken during the science verification phase of ALMA. These observations were taken using the new long baseline capability of ALMA and provide a unique view of this star-forming galaxy. We determine the structure of the background source by applying a Bayesian pixellated visibility-fitting lens modelling technique \citep*{rybak15}. A reconstruction of the emission line properties of the background source, using the lens model derived here is presented in a companion paper (Paper II). 

\section{ALMA continuum imaging}
\label{obs}

The ALMA science verification dataset of SDP.81 was taken over several observing blocks during 2014 October and November. The data were taken in Bands 4, 6 and 7 and comprised both continuum imaging datasets centred at 151, 236 and 290 GHz, and spectral line datasets centred on the CO (5-4), CO (8-7), CO (10-9) and H$_2$O (2-1) emission lines. In this letter, we only consider the datasets taken in Bands 6 and 7 since they have the highest continuum signal-to-noise ratio. 

The Band 6 and 7 continuum datasets were each taken using 2 spectral windows of 2 GHz total-bandwidth and with 128 channels in each of the linear polarisations (XX and YY). 
Other spectral windows were available, but these were used primarily for spectral line studies. A visibility averaging time of either 2 or 6~s was used. The observations were carried out with 31--36 antennas in the array and baseline lengths from $\sim$15 m to 15 km, with 10 per cent of the baselines shorter than 200~m. This gave an array that was sensitive to structures on angular scales between 19~arcsec and 16~mas (at 236~GHz). We refer to the ALMA partnership (2015) for more details about the observations and  data processing. We found that the visibility data for the final observing block of the Band 6 dataset had a larger rms noise, and so this observing block was removed from the dataset. Also, the absolute flux-density calibration of the first spectral window of the Band 6 dataset was higher by $\sim$28 percent relative to the other sub-bands, an off-set that we corrected for during our analysis. The post-reduction data quality of the Band 7 dataset was excellent. The total time on-source was 4.4 and 5.6 h for the Band 6 and 7 observations, respectively. 

The main goal of this letter is to reconstruct the surface brightness distribution of the background source from the visibility data. However, each dataset had over 10$^8$ visibilities, which is computationally expensive to analyse. To increase the speed of our analysis, we therefore averaged each spectral window to a single channel and used a visibility averaging time of 20~s, both of which have minimal bandwidth and time smearing effects on the final dataset. In addition, from our imaging tests, we found that the recovered structure of the source is highly dependent on the baseline data that are used. In Fig. \ref{fig:images} we present the imaging data for Bands 6 and 7 using a cut in the {\it uv}-data at baseline lengths of $\le$\,0.5 and $\le$\,2~M$\lambda$ (where $\lambda$ is the observing wavelength; 1.3 and 1 mm at Band 6 and 7, respectively). The $\le$\,0.5~M$\lambda$ images show that at mm-wavelengths the observed system comprises four images in a cusp configuration with evidence for a low-surface brightness Einstein ring. The $\le$\,2~M$\lambda$ image shows that the large arc is composed of several structures with varying surface brightness, and although the Einstein ring is effectively resolved out at these scales, there is still evidence of emission from this part of the source. The total flux-density of the source at Bands 6 and 7 at $\le$\,0.5~M$\lambda$ from our fit to the visibility data are 21.7\,$\pm$\,2.2 and 38.9\,$\pm$\,3.9~mJy, respectively, assuming a 10 percent uncertainty in the flux-density calibration.

For the {\it uv}-data at $\ge$\,1~M$\lambda$ (i.e. removing the shortest baselines), we found that the extended arc is mostly resolved out, which demonstrates that most of the emission at mm-wavelengths is extended; it is only the most compact components within the individual images that are detected with the $>$\,2.5~km baselines. We carried out additional tests to determine the scales on which most compact structures are resolved out. We found that they are detected up to around 5~M$\lambda$, that is, around 30~mas-scales. Based on these results, we used the visibility data at {\it uv}-distances $\le$\,2~M$\lambda$ for our lens modelling only, because there is little or no extended structure detected on baselines longer than this. Also, a cut was used instead of a {\it uv}-taper because this limited the number of visibilities needed for the gravitational lens modelling to a manageable dataset of about 1.1--1.5\,$\times$\,10$^5$ visibilities per spectral window. 

With a Briggs weighting scheme (robust $= 0$), which best represents the intrinsic weighting scheme of visibility lens modelling within a Bayesian framework \citep{Junklewitz13,rybak15}, the synthesised beam-size of the $\le$\,2~M$\lambda$ dataset was 95~$\times$~71~mas at a position angle of 64.5 deg east of north (at 236 GHz). We note that the emission line data are also only detected on the shorter baselines, and so our choice of {\it uv}-cut will allow the continuum and molecular gas emission to be compared on similar angular-scales (see Paper II). 

\section{Gravitational lens mass modelling and source reconstruction}
\label{model}

\begin{figure*}
\begin{centering}
\begin{tabular}{llll}
{\includegraphics[height=3cm, clip=true]{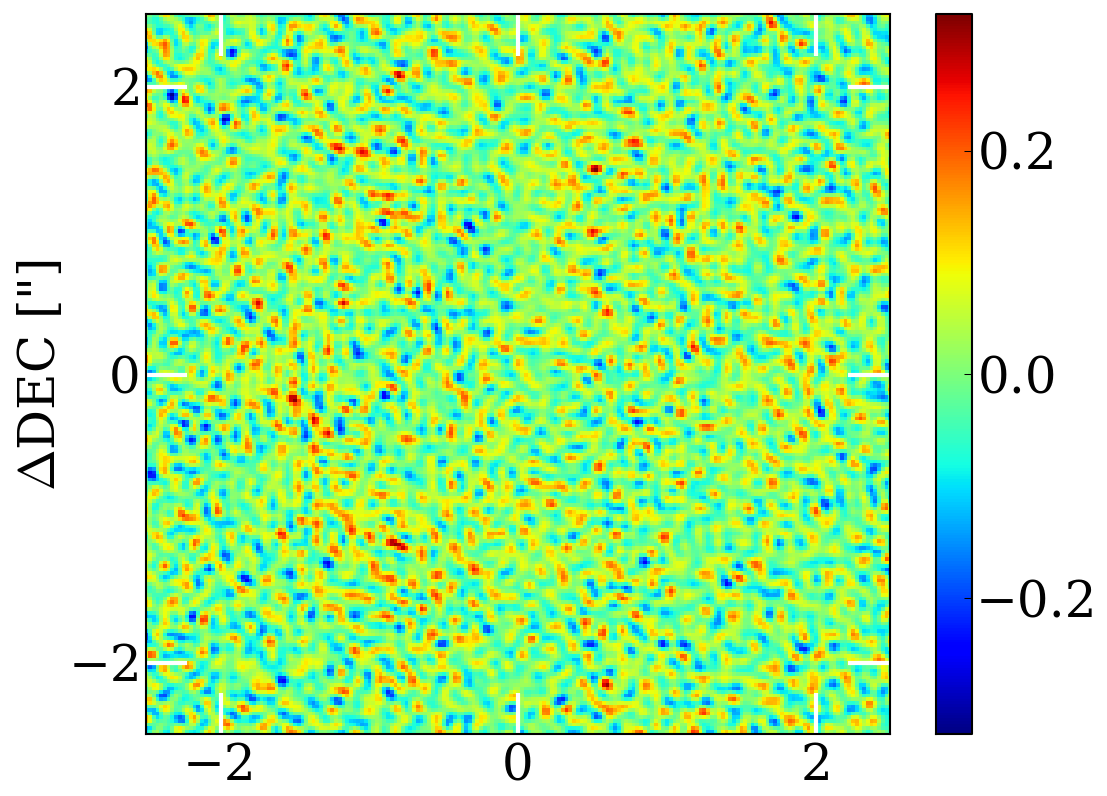}}&\hspace{-0.4cm}
{\includegraphics[height=3.08cm, clip=true]{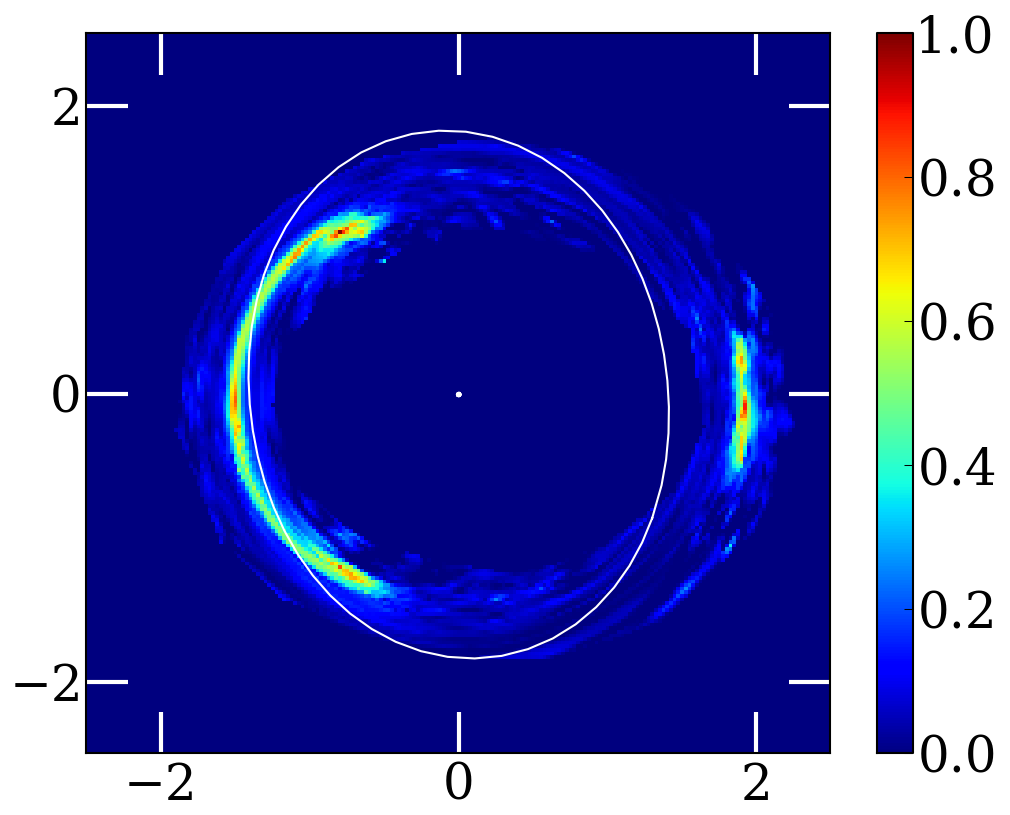}}&
{\includegraphics[height=3.08cm, clip=true]{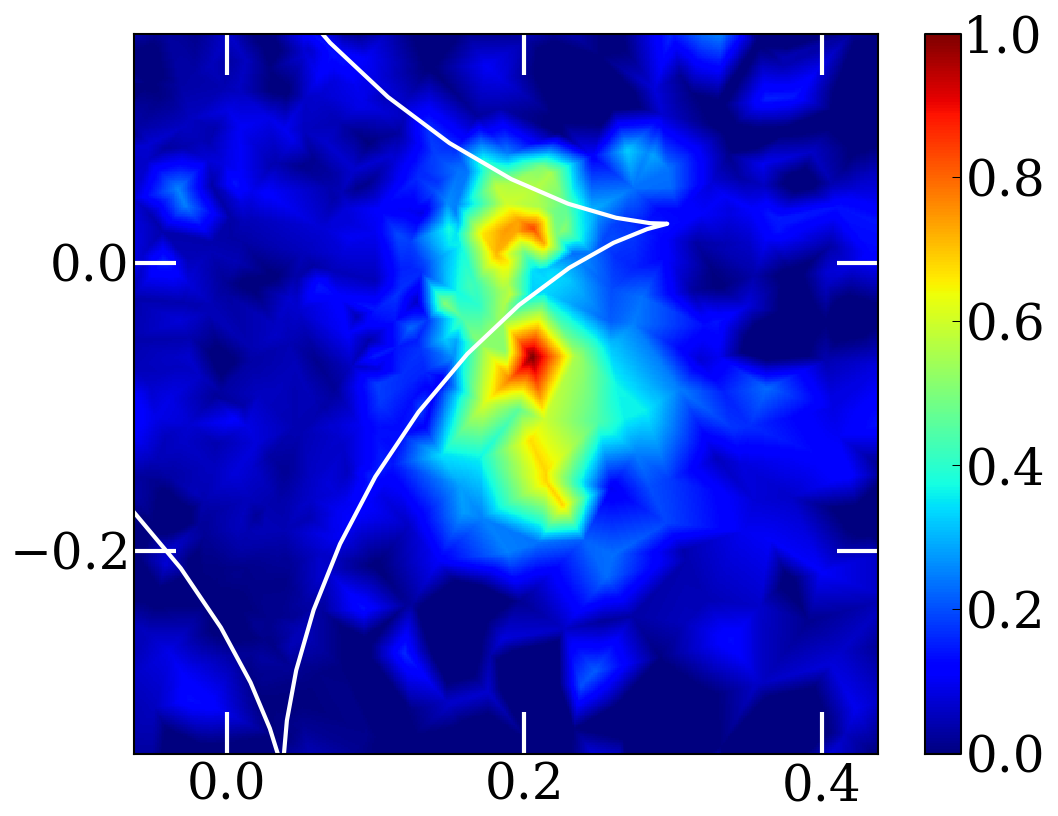}}&\hspace{-0.4cm}
{\includegraphics[height=3cm, clip=true]{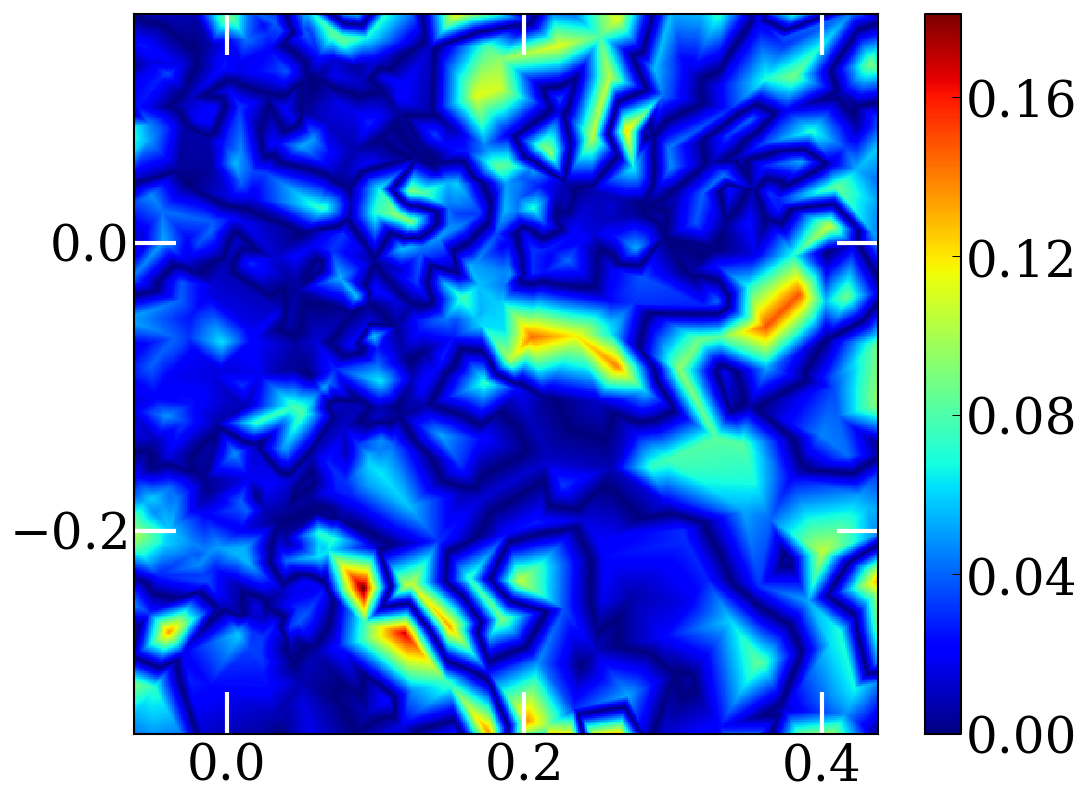}}\\
{\includegraphics[height=3.2cm, clip=true]{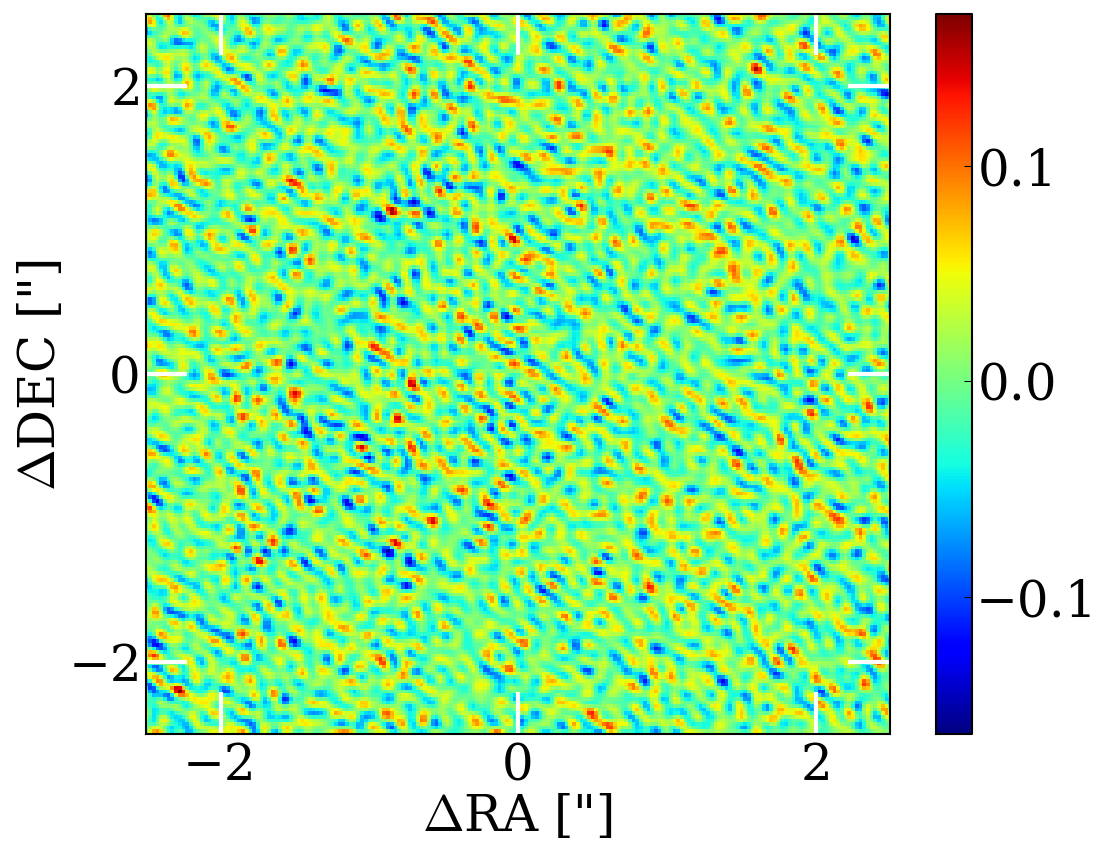}}&\hspace{-0.4cm}
{\includegraphics[height=3.29cm, clip=true]{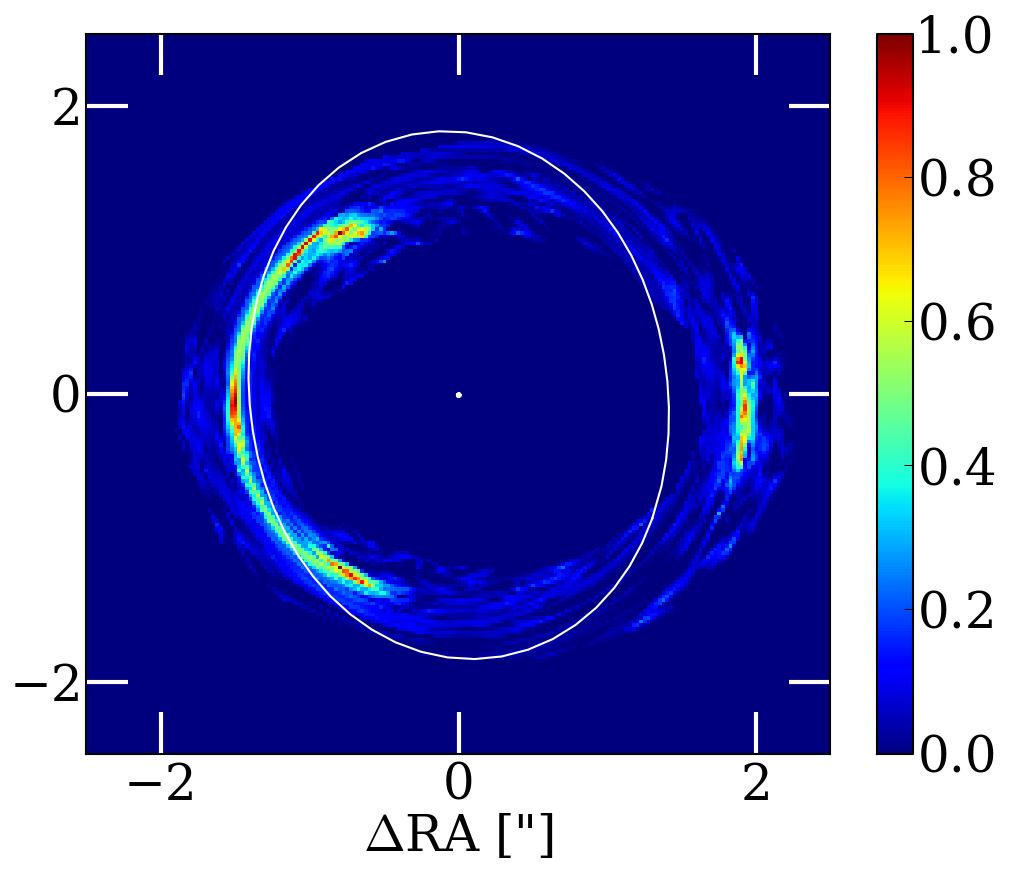}}&
{\includegraphics[height=3.29cm, clip=true]{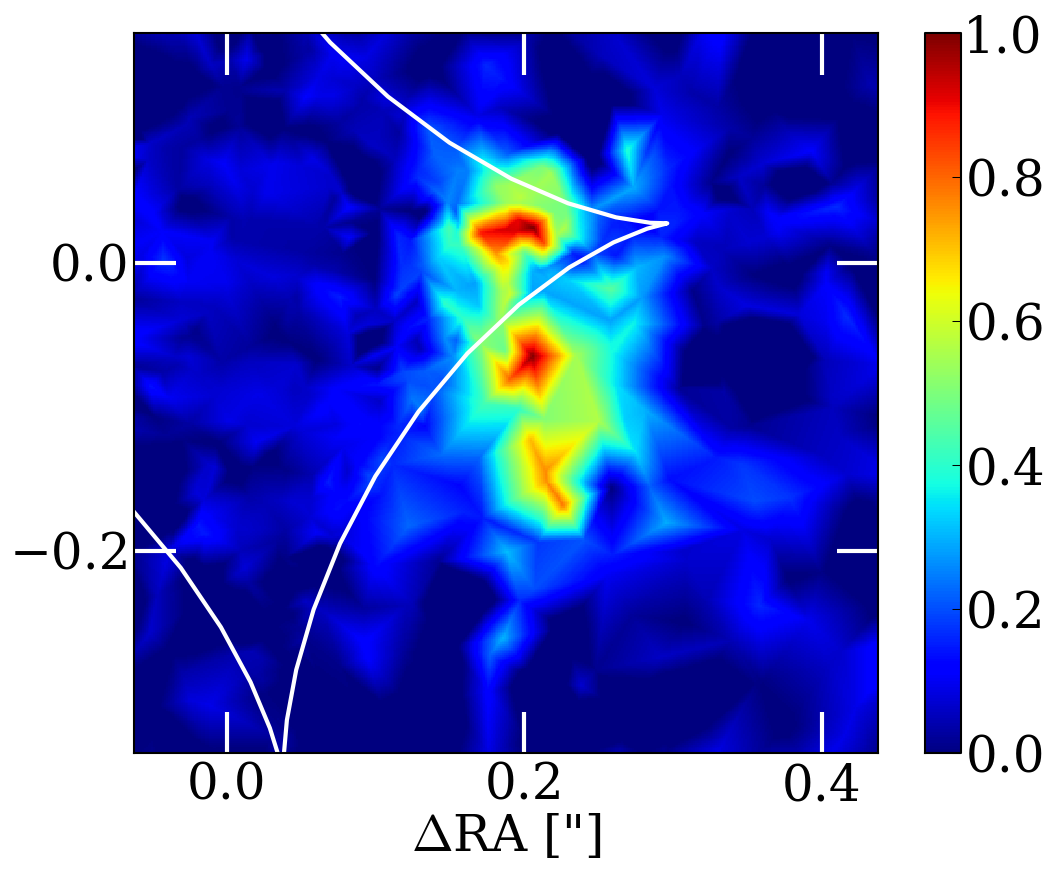}}&\hspace{-0.4cm}
{\includegraphics[height=3.2cm, clip=true]{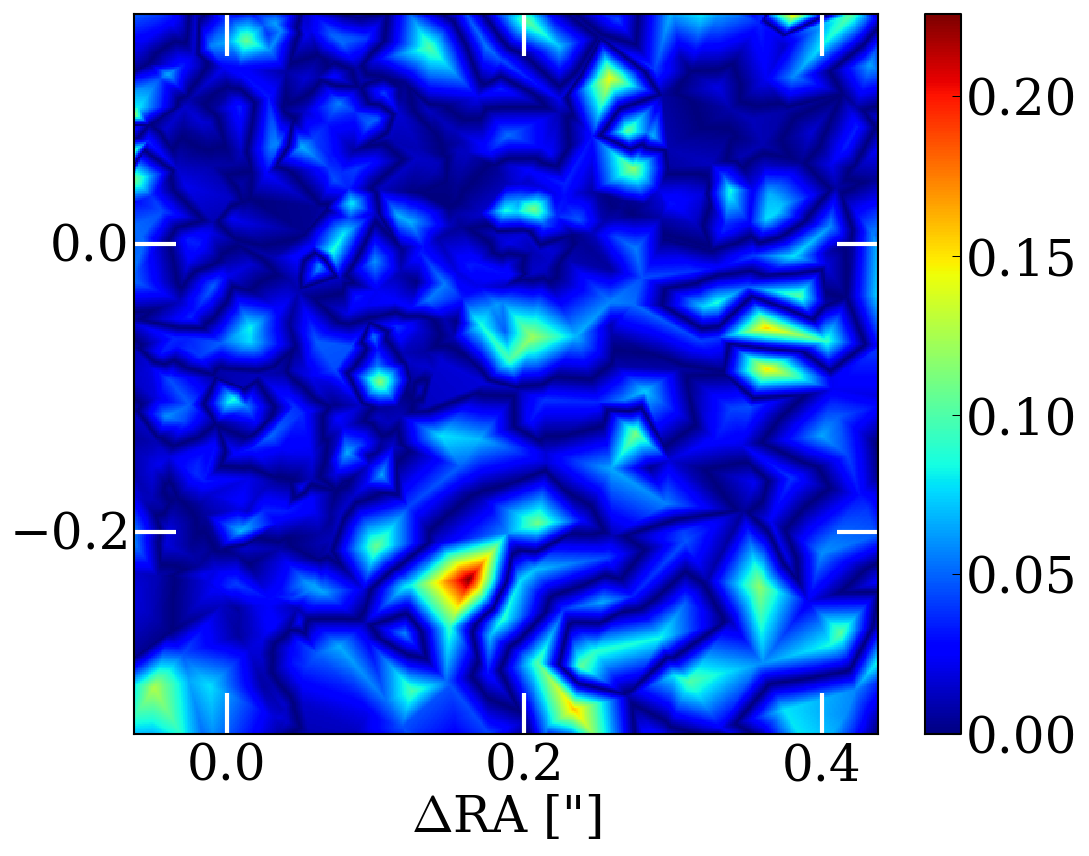}}
\end{tabular}
\caption{(Left) The image-plane residuals from the Fourier transform of the residual visibilities (data$-$model) and gridded assuming uniform weighting, (left-middle) the modelled image-plane brightness distribution with critical curves, (right-middle) the reconstructed source-plane brightness distribution with caustics, (right) and the uncertainty on the reconstructed source surface brightness distribution at Band 6 (upper) and at Band 7 (lower). The image-plane residuals are in units of mJy~beam$^{-1}$. The peak surface brightness of the reconstructed source is normalised to 1 units pixel$^{-1}$. The lens centre is at position (0, 0). The source surface brightness crosses the caustics, which results in parts being either quadruply or doubly imaged. There is also diffuse and extended structure that produces the Einstein ring emission.}
\label{fig:band6}
\end{centering}
\end{figure*}

In order to reconstruct the smooth lensing potential and the background source surface brightness distribution, we use the Bayesian visibility-fitting method of \citet*{rybak15}. This method extends the gravitational imaging technique introduced by \citet{Vegetti09} to interferometric data. 

Due to the incomplete sampling of the {\it uv}-plane, the reconstruction of the sky surface brightness distribution from the measured visibility function is a non-trivial inverse problem. Moreover, the pixel-to-pixel noise in the sky surface brightness maps is correlated, and since the individual baselines filter structure in the source on various angular-scales, the observed surface brightness (in coarsely resolved imaging) is not necessarily conserved.  As a result, fitting to the data directly in visibility space has become the established method to determine the lens model parameters and to reconstruct the surface brightness distribution of the background source from interferometric data (e.g. \citealt{Wucknitz04,Hezaveh13a,Bussmann2013}). However, these analyses have so far been dependent on assuming a parametric form, typically multiple Gaussian or Sersic profiles, for the background source. These are unlikely to represent well the complex surface brightness distribution of dusty star-forming galaxies on sub-kpc scales \citep{rybak15}. It is for this reason that we instead use a pixelated source model to reconstruct the extended emission from the high angular resolution ALMA observations. 

The technical implementation of pixelated source grids and their application to interferometric data are discussed in detail by \citet{Vegetti09} and \citet{rybak15}, respectively. To summarise, the source-plane grid is constructed using a Delaunay tessellation so that the highly magnified regions are sampled more densely than those with lower magnification, thus saving computational time. This also enables the small-scale structure in the highly magnified regions of the source to be better sampled. The modelling method optimises the Bayesian evidence by minimising a combination of the $\chi^2$ measured in the visibility space and a regularisation performed on the source plane grid. By including the regularisation term, we break the ill-posed nature of the lensing equation by enforcing a degree of smoothness in the source surface brightness distribution, while avoiding overfitting the noise. Here, we adopt the gradient rather than the curvature regularisation to prevent the source from being overly smoothed. A detailed discussion of various levels of source regularisation is given by \citet{Suyu06b}.

We use an elliptical power-law mass distribution with an external shear to describe the lensing potential. This model is defined by a mass parameter ($\kappa_0$; the normalization of the surface-mass density), an ellipticity ($q$) and position angle ($\theta$; east of north), a power law density-profile ($\gamma$), a shear ($\Gamma$) and a shear position angle ($\Gamma_\theta$; north of west). The most probable {\it a posteriori} model parameters are presented in Table \ref{tab:lens} with their uncertainties. These are consistent with the gravitational lens modelling of the SMA data by \citet{Bussmann2013} and of the {\it HST} data by \citet{Dye14} at the 2$\sigma$-level, with the small differences being attributed to the choice of the mass model, for example, Bussmann et al. do not include shear in their model. 

With the lens model in hand, we reconstruct the source surface brightness map for each of the two spectral window datasets in Bands 6 and 7 separately, and then take an average to improve the signal-to-noise ratio. This is done partly to provide a consistency check between the source reconstructions for each spectral window, but also to limit the size of the visibility datasets that are fitted. We estimate the noise level for each baseline at a given time and frequency by calculating the rms of the visibilities measured within the same observing block (typical length of approximately 30 minutes). This could lead to a biased noise estimate if the system temperature varies sharply on short time-scales, but this was not found to be the case. 

In Fig.~\ref{fig:band6} we show the average reconstructed source surface brightness distribution and the resulting sky-model for the Band 6 and 7 datasets. In addition, we also present error estimates for the reconstructed source. These are taken from the rms scatter in the source reconstructions between the different spectral windows of each band.  We find that the structure of the source surface brightness distribution is in general consistent in the two bands, which provides an independent check of our source reconstruction. However, there are variations in the relative surface brightness of the structure within each band (see below for discussion). Note that the high surface brightness inner region of the source is reconstructed at higher resolution and at a higher signal-to-noise ratio compared to the rest of the source plane.

We have determined the residuals between the data and the model in the visibility plane, and find that they are within the uncertainties of the individual visibilities in the real and imaginary parts. The residuals are also within the expected rms over the whole dataset. For example, 4.7 and 0.3 percent of the visibilities are out with the 2- and 3-$\sigma$ levels, respectively, which is in good agreement with what is expected from a Gaussian distribution (white noise). However, to illustrate the goodness of fit of our model, we also show in Fig.~\ref{fig:band6} the image plane residuals. Note that the image plane residuals are dependent on the weighting scheme used when gridding the visibilities.

\begin{table*}
\begin{center}
\caption{The best lens model parameters for SDP.81 derived from an analysis of the ALMA continuum visibility data.}
\begin{tabular}{@{}cccccccc @{}} \hline
$\kappa_0$	 (arcsec)		& $q$			& $\theta$ [deg]	& $\gamma$	& $\Gamma$	& $\Gamma_\theta$ [deg]\\ \hline
1.606$\pm$0.005			& 0.82$\pm$0.01 			& 8.3$\pm$0.4 				& 2.00$\pm$0.03 			& 0.036$\pm$0.004 		& 3.0$\pm$0.2 \\ \hline
\end{tabular}
\label{tab:lens}
\end{center}
\end{table*}

\section{The intrinsic source properties of SDP.81}
\label{disc}

The reconstructed continuum emission of SDP.81 at 236 and 290 GHz possesses a complex morphology; there are three distinct regions of the source that are resolved in our pixelated reconstructions, but there is also evidence of a wider region of low-surface brightness emission (see Fig.~\ref{fig:band6}). This outer region is consistent with the extended emission that forms the Einstein ring seen in our low resolution images (see Fig.~\ref{fig:images}). It is clear that the sub-kpc structure of the galaxy is not uniform, but contains several regions of intense dust emission that are presumably associated with recent star-formation. In particular, the central part of the source is elongated and has an extent of $\sim$1.9$\times$0.7 kpc. We interpret this as evidence of a star-forming disk (e.g. as in GN20; \citealt{hodge14}), but it could also be due to the multiple components of an ongoing merger (e.g. \citealt{Swinbank06, Tacconi2008, Engel2010}). We note that photo-dissociation region (PDR) models for SDP.81 from an analysis of the [O~{\sc iii}] lines in the far-infrared spectrum, as measured with {\it Herschel}, constrain the size of the emitting region to be ~500--700 pc (effective radius; \citealt{valtchanov11}), which is consistent with the overall size of the central star forming region that we detect. However, the individual regions with the most intense dust emission are smaller than this size, that is, $\leq$\,500~pc. 

The extent and surface brightness distribution of the reconstructed source are considerably different than reported by \citet{Bussmann2013}, who modelled the SMA data at 880~$\mu$m (340 GHz) using a single Sersic profile with an effective radius of $\sim$4.1~kpc and at a position angle that is almost perpendicular to the extension seen in the ALMA reconstruction. However, as \citet{Bussmann2013} discuss, their single parametric source model is too simple to fit the observed surface brightness distribution of the 880~$\mu$m emission at 0.6 arcsec angular resolution; this highlights the advantage of using pixelated source reconstructions and 100~mas angular resolution observations to study the intrinsic properties of gravitationally lensed starburst galaxies.

We estimate the lensing magnification $\mu$ of the source by calculating the ratio of the reconstructed emission in the source-plane and image-plane. The total magnification is found to be $\mu_{\rm tot} =$~17.6\,$\pm$\,0.4. This magnification is dominated by the central star-forming disk, which has a magnification of  $\mu_{\rm SF~disk} =$~25.2\,$\pm$\,2.6. Our total magnification is a factor of 1.6 higher than those reported by \citet{Bussmann2013} and \citet{Dye14}, which is most likely due to the differing source structure that we find compared to their studies; compact sources that are located close to the lens caustic can have very high magnifications. In addition, our analysis directly measures the magnification of the resolved continuum emission from the dust, which is most relevant when studying the star-forming properties of SDP.81. The compact structure of the source and the high magnification that we find are consistent with a potential size-bias in gravitationally lensed sources (e.g. \citealt{Tacconi2008}) and the results for this source should not necessarily be extrapolated to the general sub-mm galaxy population. 

Using our measurement of the total magnification, we have determined the intrinsic properties of SDP.81 from a fit to the UV/optical-to-sub-mm broad-band spectrum that was carried out by \citet{Negrello14}. We find a star-formation rate of 315\,$\pm$\,60~M$_{\odot}$~yr$^{-1}$, a far-infrared luminosity of 3.1\,$\pm$\,0.4~$\times$~10$^{12}$~L$_{\odot}$ and a dust mass of 6.4\,$\pm$\,1.5~$\times$~10$^{8}$~M$_{\odot}$. Here, we only consider the properties derived from the far-infrared part of the spectrum, since this is the parameter space where our magnification is valid. These results are consistent with what is typically found for sub-mm galaxies at redshift $\sim$2.5 (e.g. $\sim$400~M$_{\odot}$~yr$^{-1}$; \citealt{coppin08}).

\begin{figure}
 \begin{tabular}{ll}
 \hspace{-0.3cm}
\includegraphics[scale=0.24]{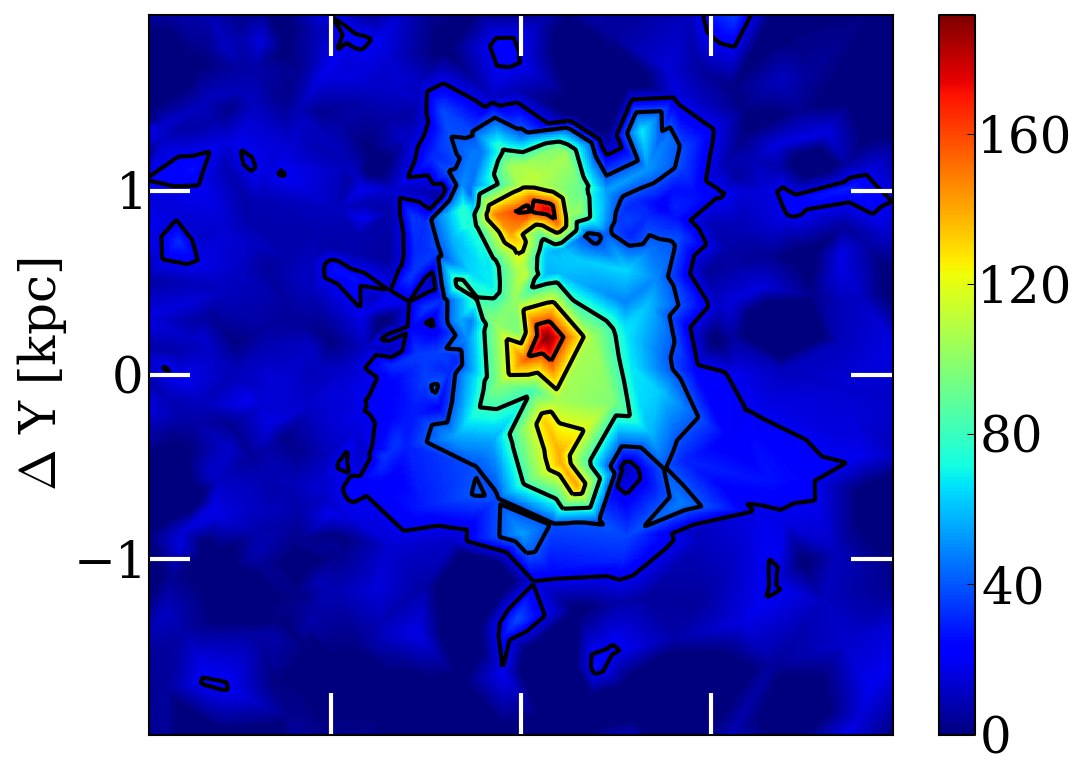}&\hspace{-0.4cm}
\includegraphics[scale=0.24]{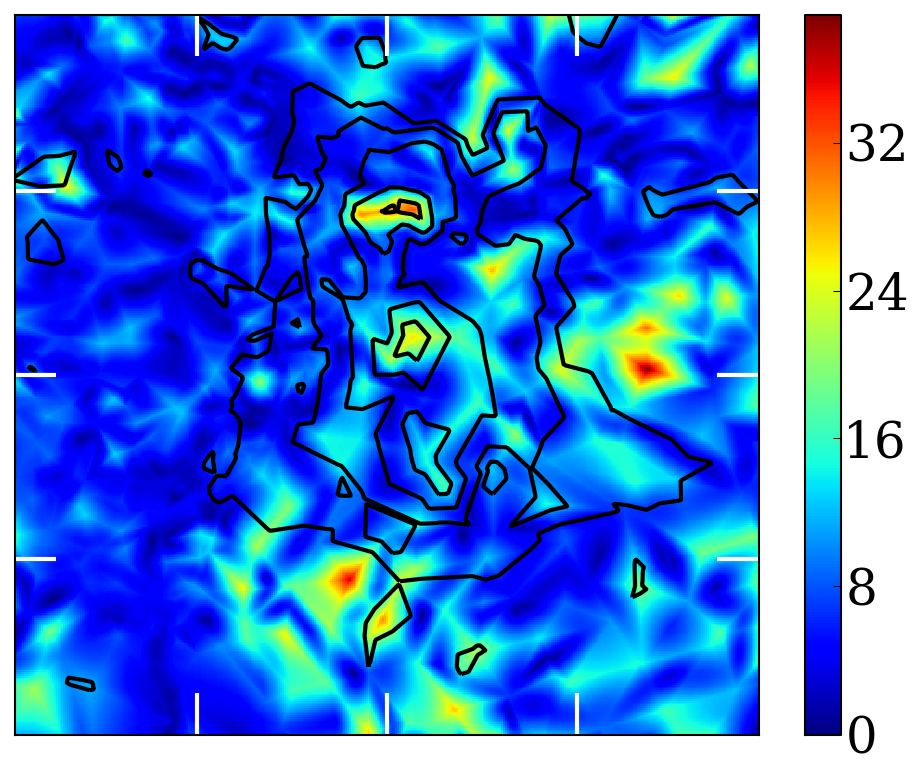}\\
 \hspace{-0.3cm}
\includegraphics[scale=0.24]{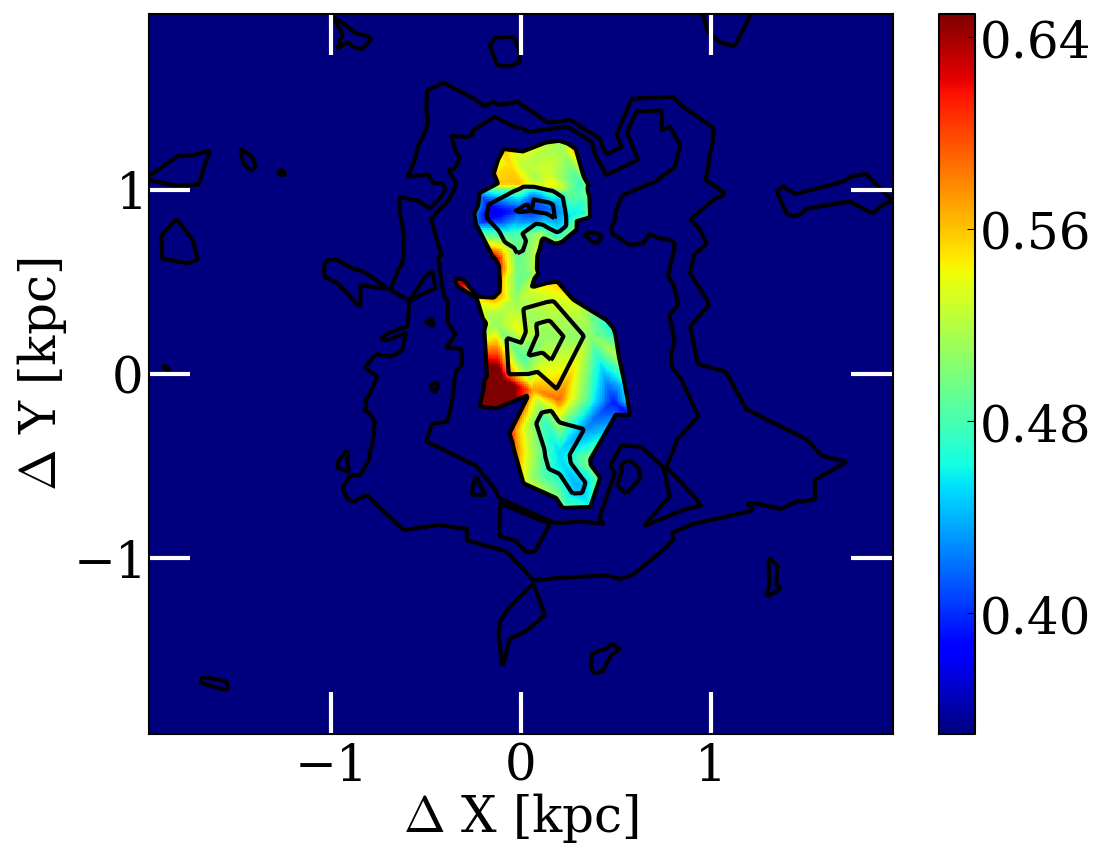}&\hspace{-0.4cm}
\includegraphics[scale=0.24]{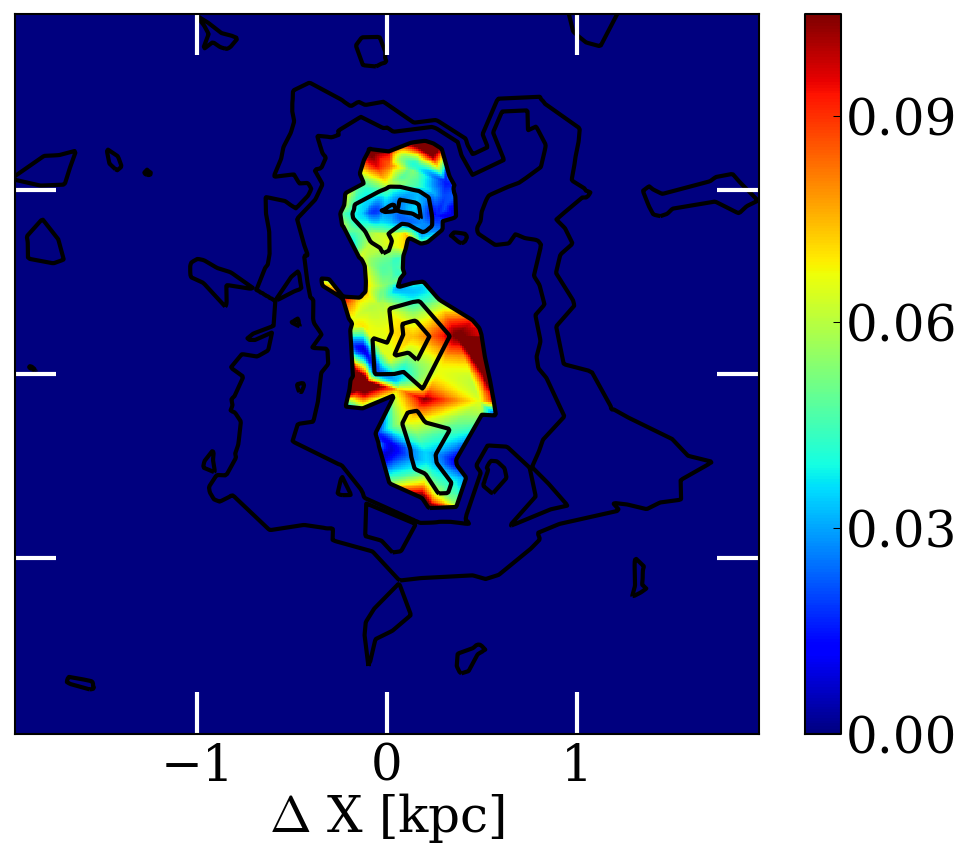}
 \end{tabular}
\caption{(Upper) The star-formation rate density (left) and the uncertainty (right) of the SDP.81 reconstructed source at redshift 3 shows multiple regions of intense star-formation. The colour-bar is in units of M$_{\odot}$~yr$^{-1}$~kpc$^{-2}$ and the peak star-formation rate density is 190\,$\pm$\,20~M$_{\odot}$~yr$^{-1}$~kpc$^{-2}$. (Lower) The flux-ratio between 236 and 290 GHz (left) and the uncertainty (right) for a region with a star-formation rate density $\ge$~80~M$_{\odot}$~yr$^{-1}$~kpc$^{-2}$. The colour-bar is in units of flux-ratio. For all plots the iso-contours of the star-formation rate density are set at 20, 40, 80, 120 and 160~M$_{\odot}$~yr$^{-1}$~kpc$^{-2}$.}
\label{fig:sfr_density}
\end{figure}

In Fig. \ref{fig:sfr_density}, we show the star-formation rate density of the reconstructed source, under the assumption that all the heated dust emission is solely due to star formation. Also shown is the uncertainty in the star-formation rate density, which was calculated from the rms over all spectral channels. We find that the extended component has a moderate star-formation rate density of $\sim$20--30~M$_{\odot}$~yr$^{-1}$~kpc$^{-2}$, but there are multiple regions of intense star-formation ($\geq$\,100~M$_{\odot}$~yr$^{-1}$~kpc$^{-2}$) as opposed to a single starburst site. The star-formation disk is forming stars at a mean rate of over 100~M$_{\odot}$~yr$^{-1}$~kpc$^{-2}$. This is several orders of magnitude higher than is seen within disk galaxies in the local Universe \citep{leroy13}, but is comparable to the extended disk seen in GN20 at redshift 4 \citep{hodge14}. The maximum star-formation rate density in our reconstructed map is 190\,$\pm$\,20~M$_{\odot}$~yr$^{-1}$~kpc$^{-2}$, which is below the theoretical expectation for Eddington-limited star-formation by a radiation-pressure supported starburst ($\sim$1000~M$_{\odot}$~yr$^{-1}$~kpc$^{-2}$). There are several other examples of high redshift sub-mm luminous galaxies that have sub-Eddington starbursts \citep{Bussmann2012,hodge14}, but this is the first time that the star-formation density of a galaxy has been directly mapped on sub-50~parsec-scales thanks to gravitational lensing and the new high resolution imaging capability of ALMA. Clearly, high resolution imaging of the larger sample of gravitational lenses found during the SPT and H-ATLAS lens surveys, coupled with pixelated source reconstructions, is needed to determine if these galaxies as a population are considerably below the Eddington-limit for radiation-pressure supported starbursts.

We also see tentative evidence for a change in the dust continuum slope over the resolved extent of the intense star-forming regions within SDP.81 (see Fig.~\ref{fig:sfr_density} for the relative brightnesses of the regions). Between 230 and 290 GHz, the flux-ratio of the entire source is 0.56\,$\pm$\,0.08 (or a spectral index of $\alpha =$ 2.8, where $S_{\nu} \propto \nu^{\alpha}$), which is as expected for the best fit modified black body model for the dust emission (temperature $T_{\rm D} =$~34\,$\pm$\,1~K and emissivity $\beta =$~1.5; \citealt{Bussmann2012}). We find that the central dominant clump of dust emission in our maps has a flux-ratio of 0.51\,$\pm$\,0.07 between 230 and 290 GHz, but the two additional clumps of intense dust emission to the north and south have flux-ratios of 0.41\,$\pm$\,0.03 and 0.45\,$\pm$\,0.05, respectively (equivalent to $\alpha \sim$ 4.3--3.9). This change in the spectral slope of the continuum emission suggests that the different sites of intense star-formation have different physical conditions, but further observations at high angular resolution with Band 9 (and 10) will be needed to determine the extent of the variation in the dust temperature and/or optical depth at these locations. However, our results show that global temperature models are likely too simple to explain the emission from galaxies with multiple sites of star-formation, as expected.

\section{Summary}
\label{conc}

We have presented new continuum imaging of the gravitationally lensed starburst galaxy SDP.81 at redshift 3.042 using the new long baseline capability of ALMA. We find that on $\sim$0.5~arcsec-scales, the background source at mm-wavelengths is gravitationally lensed into four images in a standard cusp configuration with a low-surface brightness Einstein ring. On $\sim$100~mas-scales, the background source is resolved into two gravitational arcs that clearly show multiple regions of enhanced dust emission. Most of the source is completely resolved out on 30-mas-scales, which corresponds to baseline lengths of $\sim$5~M$\lambda$. Our pixelated source reconstruction from the visibility data, shows that the unlensed source is $\sim$1.9~kpc in diameter and elongated; consistent with a star-forming disk that has multiple regions of intense star-formation. This small linear size of the source is consistent with the expected size-bias associated with gravitationally lensed starbursting galaxies. The star-formation rate density varies from $\sim$20 to 190~M$_{\odot}$~yr$^{-1}$~kpc$^{-2}$ across the source, which is as expected for a sub-Eddington-limited starburst. We also detect a tentative variation in the far-infrared continuum slope across the source. 

The results presented in this letter represent the first direct mapping of the star-formation density of a galaxy at such small scales and highlight the power of combining gravitational lensing and the high resolution imaging capability of ALMA.

\section*{Acknowledgments}
This paper makes use of the following ALMA data: ADS/JAO.ALMA\#2011.0.00016.SV. ALMA is a partnership of ESO (representing its member states), NSF (USA) and NINS (Japan), together with NRC (Canada), NSC and ASIAA (Taiwan), and KASI (Republic of Korea), in cooperation with the Republic of Chile. The Joint ALMA Observatory is operated by ESO, AUI/NRAO and NAOJ.

\bsp

\label{lastpage}

\end{document}